\documentclass[prb,twocolumn,showpacs,showkeys]{revtex4}%
\usepackage{graphicx}
\usepackage{times}                 
\usepackage{amsmath}
\usepackage{amsfonts}
\usepackage{amssymb}
\usepackage{times}
\date{\today}

\newcommand{\Heb}{H_{eb}}
\newcommand{\Hc}{H_{c}}

\newcommand{\Jfm}{J_{FM}}

\newcommand{\Jint}{J_{int}}

\newcommand{\Mshif}{M_{AF,int}}
\newcommand{\Rc}{R_c}
\newcommand{\Lc}{L_c}
\newcommand{\tsh}{t_{sh}}

\begin{document}
	\title[Exchange bias effect in core-shell nanowires]
	{Exchange bias effect in cylindrical nanowires \\ with ferromagnetic core and polycrystalline antiferromagnetic shell}
	\author{A.~Patsopoulos}
	\email{apats@phys.uoa.gr}
	\affiliation{Department of Physics, National and Kapodistrian University of Athens, Athens, GR-15784}
	\author{D.~Kechrakos}
	\email{dkehrakos@aspete.gr}
	\affiliation{Department of Education, School of Pedagogical and Technological Education, Athens, GR-14121 }
	
	\keywords{exchange bias; core-shell nanowires; shape anisotropy; domain wall; Monte Carlo; Stoner-Wohlfarth}
	\pacs{75.60.Jk; 75.75.Jn; 75.75.Fk; 75.78.Fg }		
\begin{abstract}
We model the exchange bias effect in thin cylindrical nanowires composed of a ferromagnetic core and an  antiferromagnetic shell implementing a classical spin Hamiltonian and Monte Carlo simulations. 
We address systematically the effect of shell polycrystallinity on the characteristic fields of the isothermal hysteresis loop (coercivity, exchange-bias)  and their angular dependence upon the direction of the applied / cooling field. 
We relate the observed trends to modifications of the underlying magnetization reversal mechanism. 
We fit our simulation results to an extended Stoner-Wohlfarth model with  effective off-axis unidirectional anisotropy and demonstrate that shell polycrystallinity could lead to maximum exchange bias effect in an off-axis direction.
Our results are in qualitative agreement with recent experimental studies of Co/CoO nanowires. 
\end{abstract}
\maketitle
%================================================================
\section{Introduction}
%================================================================
Tailoring the magnetic properties at the nanoscale has attracted great research interest in recent years, due to the plethora of prospective applications\cite{sko03}. 
Advances in template-assisted electrodeposition\cite{fer99} enhanced substantially the research effort in  quasi-one dimensional nanostructures, such as nanowires and nanotubes. 
These systems are characterized by enhanced shape anisotropy, which renders them prominent candidate materials for advanced technologies ranging from magnetic recording\cite{sel01} to biomedicine\cite{bau04,tar06}.
On the other hand, the quasi-one dimensional geometry is ideal for efficient manipulation of magnetic solitonic excitations, such as domain walls, with potential applications to 3D magnetic memory\cite{par08}. 
The magnetization reversal process lies in the heart of research related to magnetic nanowires. 
Their quasi one-dimensional shape leads to a complex reversal mechanism, consisted of domain wall (DW) nucleation, propagation and annihilation\cite{fer99,sel01,thi06}. 
Furthermore, it has been demonstrated that the magnetization reversal mechanism and concomitant anisotropy can be tailored by alloying \cite{bra15}, by periodic chemical modulation as in multisegmented nanowires \cite{ber17} and by morphological modulations, as in diamater-modulated nanowires\cite{pal15}.

In an alternative route to tailoring of the magnetic anisotropy of magnetic nanostructures, the exchange bias (EB) effect\cite{mei56,mei57} has been extensively studied in nanostructures with a ferromagnetic (FM) core - antiferromagnetic (AF) shell  morphology\cite{nog05,igl08}. 
Hysteresis loops of such systems exhibit a characteristic shift after been field-cooled, usually accompanied by magnetic hardening. Materials presenting these properties are currently implemented in spintronics applications, like spin-valves and magnetic tunnel junctions\cite{koo96}.
Elongated nanostructures such as nanowires\cite{mau09,tri10,gan17} and nanotubes\cite{buc15,pro13} with FM core - AF shell morphology have also been experimentally investigated showing exchange-bias and the accompanying effects. 
Maurer \textit{et al}\cite{mau09} compared the hysteresis properties of cylindrical Co and Co/CoO nanowires and demonstrated the suppression of the coercive field due to surface oxidation as well as an anomalous temperature dependence, which was attributed to the thermal fluctuations of the oxide shell.
In a search for optimum applied field direction for enhancement of the EB effect, Tripathy \textit{et al}\cite{tri10} investigated the angular dependence of the EB field of lithographically grown Co/CoO nanowires and demonstrated the increase of the EB field in off-axis directions due to competing unidirectional and uniaxial (shape) anisotropies.
Gandha \textit{et al}\cite{gan17} reported giant EB effect in  electodeposited Co/CoO nanowires with monocrystalline core and polycrystalline shell and enhancement of the EB effect in an off-axis direction relative to the nanowire axis.

Due to the crucial role played by the FM-AF interface spin structure in the EB effect\cite{nog05} a realistic description of the structure and dynamics of the AF layer is necessary. 
In a previous work\cite{pat17}, we studied uniaxial FM core/AF shell nanowires in order to study the impact of the AF shell on the loop characteristics and in the magnetization reversal mechanism. We showed that the interface exchange coupling produces a weak exchange-bias field and a suppression of coercivity relative to the bare FM nanowire. 
This behavior was attributed to the presence of unsatisfied FM-AF interface bonds that act as a sequence of nucleation centers leading to a secondary reversal mechanism that acts in synergy to domain wall propagation and eventually to mobility enhancement. 
However, Co/CoO nanowires prepared by surface oxidation of electodeposited Co nanowires develop a polycrystalline shell\cite{gan17}. 

The effect of shell granularity in FM/AF bilayers has been recently studied experimentally\cite{cha17} for Co/CoO bilayers and it was shown to cause enhancement of the EB field. 
However, to the best of our knowledge, a study of the effect of shell granularity on the the coercivity, the EB field and their angular dependence in the case of FM-AF core-shell nanowires has not been addressed yet.  
  
In the present work, we implement the Monte Carlo method to study the effects of shell granularity on the EB behavior of cylindrical nanowires composed of a monocrystalline Co core and polycrystalline CoO shell. 
Shell granularity is shown to soften the magnetic anisotropy of the shell leading to enhancement of coercivity and suppression of the EB field. 
Additionally, shell granularity is shown to produce a monotonous drop of the coercive field and a non-monotonous drop of the EB field. 
The appearance of an optimum EB value is interpreted as an effective off-axis uniaxial anisotropy in the framework of a mesoscopic Stoner-Wohlfarth (SW) model. 
Finally, we compare our model predictions to recently reported measurements of the EB effect in Co/CoO nanowires\cite{gan17}. 

%================================================================
\section{Model and Simulation Method}
\begin{figure}[htb!]
\centering
\includegraphics[width=0.95\linewidth]{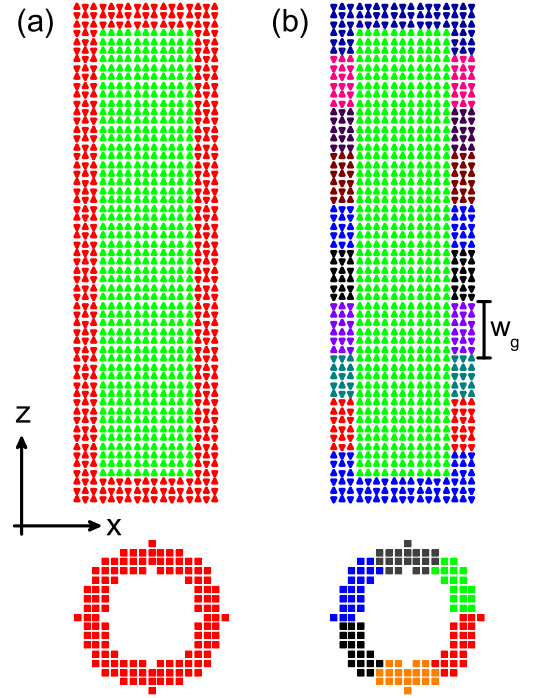}
\caption{ 
(Color online) Cutting planes (top) and shell cross sections (bottom) of cylindrical core/shell nanowires with $R_c=5a$, $L_c=50a$ and $\tsh=3a$. The nanowire shell is 
(a) monocrystalline and
(b) polycrystalline with $N_g=60~(N_z=10, N_{\phi}=6)$ grains of width $w_g\approx 5.6a$.
}
\label{fig:sites2}
\end{figure}

%================================================================
Nanowires are generated by cutting a cylinder along the z-axis with radius $R$ and length $L$ from an auxiliary simple cubic lattice with constant $a$. 
For the core-shell morphology we define an internal homoaxial cylinder with radius $R_c = R-t$ and length $L_c=L-t$ , where $t$ is the shell thickness. 
We approximate the shell polycrystallinity (granularity) by dividing the shell in $N_z$ cylindrical slices along the z-axis and each slice in $N_\phi$ circular sectors. 
Thus, the shell is divided in $N_g=N_zN_\phi$ identical crystallites (grains) with width $w_g=L/N_z$, as seen in Fig.~\ref{fig:sites2}.
The total energy of the magnetic system is $E=\sum_i E_i$, where the single-site energy term reads
%----------------------------------
% EQN - On-site energy
%----------------------------------
\begin{eqnarray}
E_i = 
-\frac{1}{2} \widehat{S}_i \cdot \sum_{<j>} J_{ij}  \widehat{S}_j 
-K_i (\widehat{S}_i \cdot \widehat{e}_i )^2  \nonumber \\
-H (\widehat{S}_i \cdot \widehat{H})
-\frac{1}{2} g \widehat{S}_i \cdot \sum_{j}  \textit{\textbf{D}}_{ij} \cdot \widehat{S}_j. 
\label{eq:energy}
\end{eqnarray}
%----------------------------------
% FIG 1 - SITES
%----------------------------------
In Eq.~\ref{eq:energy}, hats indicate unit vectors and bold symbols $3\times 3$ matrices in Cartesian coordinates. 
The $1/2$ prefactor in the first and fourth terms in Eq.~\ref{eq:energy} accounts for the double-counting of the energy contribution from each site-pair (bond).
The first term in Eq.~\ref{eq:energy} is the exchange energy between first nearest neighbors (1nn) sites. 
The exchange constant  $J_{ij}$ takes the values $J_{FM}$, $J_{AF}$ and $J_{int}$ depending on whether sites $i$ and $j$ belong to the FM, the AF or the interface region, respectively. 
The latter contains the sites of the core (shell) having exchange bonds with sites in the shell (core). 
For 1nn exchange couplings, the interface region has width $2a$ and consists of the core-interface layer and the shell-interface layer.
The second term in Eq.~\ref{eq:energy} is the uniaxial anisotropy energy.
The easy axes $\widehat{e_i}$ of the core sites are taken along the cylinder axis. 
Shell sites belonging  to the same grain have a common easy axis, which, however, varies at random between different grains. 
The anisotropy constant $K_i$ takes the values $K_{FM}$ and $K_{AF}$ depending on the location of site $i$.    
The third term in Eq.~\ref{eq:energy} is the Zeeman energy due to the applied field $H$ and the fourth term is the dipolar energy with strength $g$.
For computational efficiency, the dipolar energy term is treated in an embedded cluster approximation with a cluster radius $r_0=3a$\cite{pat17}.

To observe the EB effect the nanowires are field-cooled (FC) from a high temperature $(T>>T_c)$ to a low temperature $(T<<T_N)$ under a field well below the saturation value of the AF phase $(H_{cool} << H_{sat})$. 
At the end of the FC process the external field is swept 
$( -H_{cool} \le H \le + H_{cool} )$ at a constant rate to obtain the isothermal hysteresis loop.
The effective coercivity of the system is then defined as 
$\Hc= |H_{c1} -H_{c2}|/2$
and the EB field as
$\Heb= |H_{c1} + H_{c2}|/2$, where $H_{c1}$ and $H_{c2}$ are the coercive fields corresponding to the descending (ie. against the cooling field direction) and the ascending branch of the loop, respectively. 

The FC process is simulated using the Metropolis Monte Carlo algorithm with single-spin updates and a temperature-dependent spin-cone aperture that accelerates the approach to equilibrium\cite{new99} .
The isothermal hysteresis is simulated using the same algorithm, but with a fixed spin-cone aperture ($\theta_s\approx 3^\circ$) % at $T<<T_N$). 
Thermal relaxation at each $(H,T)$-point is done with $5\cdot 10^3$ Monte Carlo steps per spin (MCSS) for thermalization followed by $10^4$ MCSS for calculations of thermodynamic quantities. 
The latter are calculated every $10$ MCSS to minimize statistical correlations of the sampling points. 
The results at each $(H,T)$-point are averaged over $20-40$ independent relaxation sequences.
For systems with structural disorder, as the nanowires with granular shells, an average over $30-50$ samples with different realizations of disorder is performed.

In Eq.~\ref{eq:energy} we use dimensionless energy parameters
scaled  by $\Jfm$, which is arbitrarily taken as $\Jfm=10$, and 
$J_{AF}/\Jfm$ = -0.5,~ 
$J_{int}/\Jfm$ = -0.5,~ 
$K_{FM}/\Jfm$ = 0.1,~ 
$K_{AF}/\Jfm$ = 1.0 and 
$g=0.05/\Jfm$.
These parameters capture the main features of the Co/CoO exchange coupled system as previous studies of oxide-coated cobalt nanoparticles \cite{igl08,dim15} and nanowires\cite{pat17} have shown.
Temperature $T$ and the magnetic field strength $H$ are measured in units of $J_{FM}$.
Field-cooling is performed from high temperature $T_H=2.00 \Jfm$ to low temperature $T_L=0.01\Jfm$ under an applied field $H=4.0\Jfm$ with constant cooling rate $r_T=10^{-5} \Jfm/$MCSS.
The field sweep rate is also kept constant at $r_H=10^{-5} \Jfm/$MCSS to exclude variation of results with sampling time. 
%================================================================
\section{Results and Discussion}
\subsection{Isothermal hysteresis loops}
%================================================================
%----------------------------------
% FIG 2 - M(H)
%----------------------------------
\begin{figure} [htb!]
	\centering
	\includegraphics[width=0.95\linewidth]{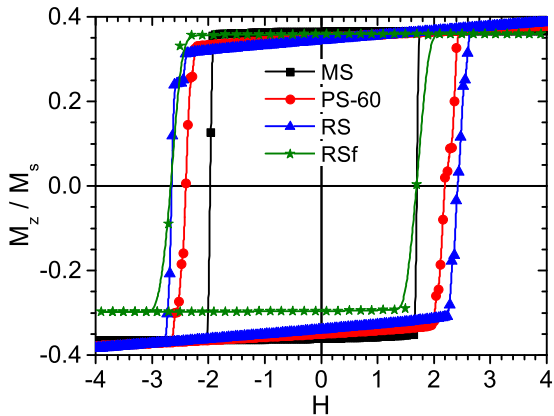}
	\caption{ 
		(Color online) Isothermal hysteresis loops of FM/AF nanowires of the same nominal size and different number of shell grains. 
		Squares (black): monocrystalline shell ($N_g=1$). 
		Circles (red): polycrystalline shell ($N_g=60$). 
		Triangles (blue): random shell ($N_g=7098$). 
		Stars (green): random frozen  shell. 
		Other parameters as in Table~\ref{table:hchb}.
	}
	\label{fig:Loops}
\end{figure}
We study first the macroscopic magnetic behavior of the nanowires, by calculating the low-temperature hysteresis loops and the characteristic fields of the loop, namely the coercivity ($\Hc$) and the exchange-bias ($\Heb$), when the cooling and reversing fields are applied along the cylinder axis (z-axis).
Different structural models of the AF shell are considered, as summarized in Table~\ref{table:hchb}.
In the limiting case of a monocrystalline shell (MS)  a single shell grain $(N_g=1)$ exists, while in the other extreme case of a random shell (RS) each shell grain contains a single site $(N_g=7098)$. 
In the intermediate case of a polycrystalline shell (PS-60), the shell contains $(N_g=60)$ grains. 
Correspondingly, the size of the shell grains is maximum in the monocrystalline sample and minimum in the random sample.  

In Fig.~\ref{fig:Loops} we compare the loops of nanowires with the same length, diameter and shell thickness, but different degree of shell crystallinity.
The loop of a random shell with frozen spins (RSf) during field sweep is also included.
A clear horizontal shift of the loops is observed in all samples as the outcome of the FC process.
An overall feature seen is the gradual shearing of the loop as the size of the grains decreases, which is related to the fact that in nanowires with many grains and many easy axes in the shell, there exists a wider distribution of energy barriers to the reversal of the core interface moments.
Furthermore, it becomes clear by simple inspection of Fig.~\ref{fig:Loops} that the loops widen while their shift decreases with decreasing grain size, which directly implies an increase of coercivity and decrease of the exchange-bias field in nanowires with small grains. 
Our numerical results for $\Hc$ and $\Heb$ for different structural models used in the present work are summarized in Table~\ref{table:hchb}.
%----------------------------------
% TABLE 1 - Hc_Heb_miu VALUES
%----------------------------------
\begin{table} %[Ht]   
	\caption{Structural parameters and results for core-shell nanowires with $\Rc=5a$, $\Lc=50a$, $\tsh=3a$.}
	\begin{ruledtabular}
		\begin{tabular}{l c c c c l}         %l=left, c=center
			System &$w_g/a$ &$\Hc$ &$\Heb$ &$\Mshif$ &$\mu$ \\
			\hline
			MS     &56.0 &1.82  &-0.14  &0.015  &0.0140 \\
			PS-60  &5.6  &2.30  &-0.11  &0.052  &0.0120 \\
			RS     &1.0  &2.54  &-0.10  &0.054  &0.0085 \\
			RSf    &1.0  &2.18  &-0.49  &0.054  &0.0099 \\
		\end{tabular}
	\end{ruledtabular}
	$w_g$=shell grain size;
	$\Mshif$=shell-interface magnetization (per spin) at the FC state;
	$\mu$=domain wall mobility (a/MCSS);
	\label{table:hchb}
\end{table}

The physical origin of the observed dependence of $\Hc$ and $\Heb$ on the grain size can be attributed to two distinct physical factors, namely the response of the shell-interface magnetization ($\Mshif$) to the applied field and the actual value of $\Mshif$ at the FC state.

\begin{figure} [htb!]
	\centering
	\includegraphics[width=0.95\linewidth]{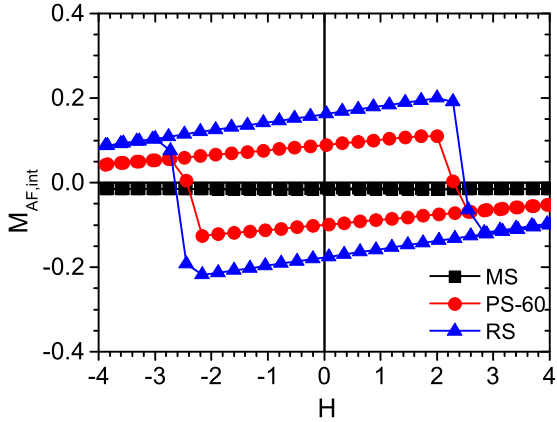}
	\caption{ 
		(Color online) Isothermal hysteresis loop of the shell interface layer, showing enhancement of AF drag effect with decreasing shell grain size. Structural parameters as in Fig.\ref{fig:Loops} 
		}
	\label{fig:loops_M3}
\end{figure}

To explain this point further, we show first in Fig.~\ref{fig:loops_M3} the magnetization of the shell-interface layer as the applied field is swept.
The observed expansion of the hysteresis loop with decreasing grain size indicates that the shell-interface spins are dragged by the core spins, via their mutual exchange coupling ($\Jint$). 
The magnetization drag mechanism becomes more efficient as the size of the grains decreases.
%----------------------------------
% FIG 3 - M3(H)
%----------------------------------
%----------------------------------
% FIG 4 - M3_Crywidth
%----------------------------------
\begin{figure} [htb!]
	\centering	\includegraphics[width=0.95\linewidth]{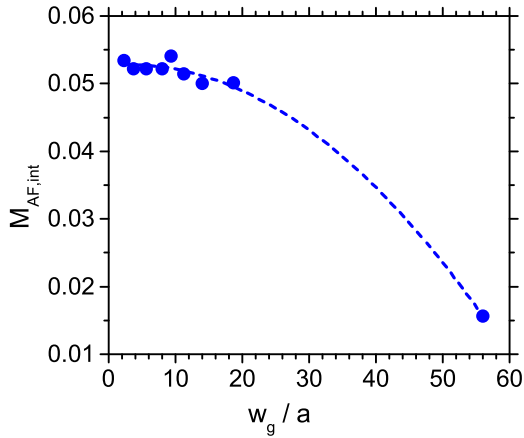}
	\caption{
		Dependence of the shell interface magnetization at the FC state on the shell grain size. Dashed line is a guide to the eye. Structural parameters as in Table~\ref{table:hchb}.
	}
	\label{fig:M3_CryWidth}
\end{figure}
The exclusive impact of the drag mechanism on $\Hc$ and $\Heb$ can be deduced by comparison of the RS and the RSf nanowires that have identical AF spin structures at the FC state (i.e. $\Mshif$ values), but for the latter the shell moments are held frozen in their FC directions  during the field sweep. 
As seen in Table~\ref{table:hchb} for these nanowires, $\Hc$ is suppressed  and $\Heb$ is dramatically  enhanced when the  magnetization drag is switched off.

Second, the role of $\Mshif$ is revealed if one compares the MS and RSf nanowires, that have very different FC states (see Fig.~\ref{fig:M3_CryWidth}), however, for both these systems the shell-interface moments remain frozen during field sweep. 
In the case of the MS nanowire, freezing of the shell-interface moments is dictated by the strong uniaxial anisotropy of the AF shell, as can be deduced from the insensitivity of shell magnetization to the applied field (Fig.~\ref{fig:loops_M3}). 
For RSf nanowires the shell moments are kept frozen during the simulation.
The relative data in Table~\ref{table:hchb} show increased value of $\Mshif$ for the RSf system which is accompanied by enhancement of both $\Hc$ and $\Heb$. 
We mention that this result is in accordance to the predictions of the Meiklejohn-Bean model that the bias-field varies linearly with the AF moment\cite{nog99}.
Overall, we conclude that the AF magnetization drag effect and the net interface moment of the AF shell $\Mshif$ act in synergy to enhance $\Hc$, while they act in competition to suppress $\Heb$.
%----------------------------------
% FIG 5 - HcHeb_Crywidth
%----------------------------------
\begin{figure} [htb!]
	\centering	
	\includegraphics[width=0.95\linewidth]{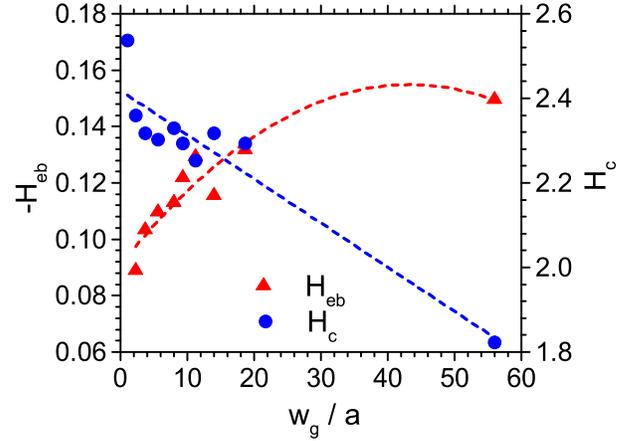}
	\caption{
		(Color online) Dependence of coercivity and exchange bias field on shell grain size. Dashed lines are guides to the eye.
		Structural parameters as in Table~\ref{table:hchb}.
	}
	\label{fig:HcHeb_CryWidth}
\end{figure}

In Fig.~\ref{fig:HcHeb_CryWidth} we show the dependence of the characteristic fields on grain size. 
In all cases the central angle of the grains is kept constant $(\phi\approx 60^{\circ})$ and only the width of the grains varies. 
We consider this is a reasonable approximation, because the nanowires studied here are thin and support transverse walls with almost coherent in-plane spin structure\cite{pat17}. 
Thus, further reduction of the grain angle, leading to increase of $N_\phi$ values, did not modify our results.
A systematic trend is seen in all cases, namely as the grain size is reduced, $\Hc$ increases and $\Heb$ decreases. This trend can be understood in the framework of the random anisotropy model\cite{her90}, where reduction of the grain size leads to an effective anisotropy of the AF shell that is an average value over several grains within the range of the exchange correlation length and thus reduced in magnitude. 
The gradual magnetic "softening" of the AF shell with decreasing grain size, enhances the magnetization drag of the interface moments and cases the observed behavior of $\Hc$ and $\Heb$. 
%================================================================
	\subsection{Magnetization reversal mechanism}
\begin{figure}[htb!]
	\centering
	\includegraphics[width=0.95\linewidth]{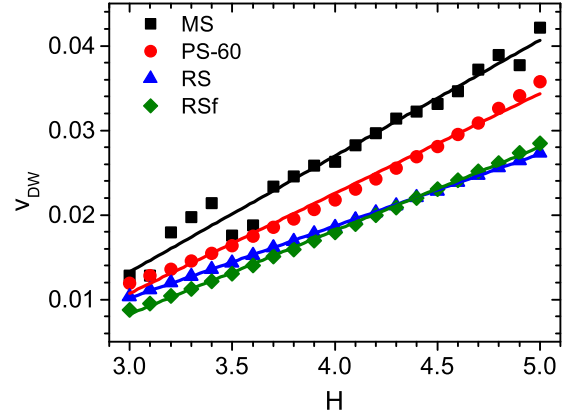}
	\caption{
		(Color online) Field-dependence of DW velocity in nanowires with different shell granularity.  
		Straight lines are linear fit to the data showing a drop of the DW mobility (slope) with increasing shell granularity. Mobility values are reported in Table~\ref{table:hchb}.
		Structural parameters as in Table~\ref{table:hchb}.
	} 	
	\label{fig:velocities}
\end{figure}

%================================================================
We discuss next, the impact of shell polycrystallinity on the underlying magnetization reversal mechanism.
%----------------------------------
% FIG 6 - M(z;t) PROFILES
%----------------------------------
\begin{figure} [htb!]
	\centering
	\includegraphics[width=0.95\linewidth]{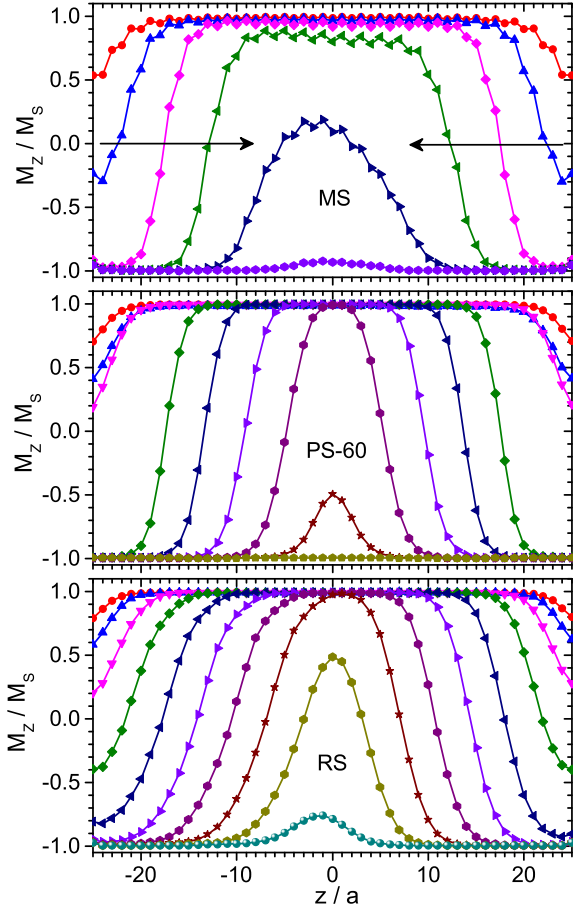}
	\caption{
		(Color online) Time-evolution of magnetization profile under application of a reverse field $H=-4.0$ of nanowires with different shell granularity. Snapshots are taken every $\Delta$t=200 MCSS starting at $t_0$=200 MCSS. Horizontal arrows indicate the propagation direction of the domain walls.
		Structural parameters as in Table~\ref{table:hchb}.
	}
	\label{fig:Mprofs3}
\end{figure}
It is well established that magnetization reversal in FM nanowires proceeds by propagation and annihilation of a pair of domain walls that nucleate at the two free ends of the nanowire \cite{fer99,hin00a,thi02}. 
The coupling to an AF shell modifies the reversal mechanism, as previous experimental\cite{mau09} and numerical works have demonstrated\cite{mau09,pat17}.
In FM/AF nanowires with a monocrystalline shell, the unsatisfied bonds at the interface act as nucleation centers of a secondary magnetization reversal mechanism, which in synergy to the domain wall propagation accelerate the reversal of the core magnetization\cite{pat17}. 
This behavior is seen in Fig.~\ref{fig:Mprofs3}a as a lowering of the core magnetization in the central region between the two domain walls.
As one can readily observe in Fig.~\ref{fig:Mprofs3}, this secondary mechanism is absent in the PS and the RS nanowires, where reversal proceeds by clear domain wall propagation. 
The reason for disappearance of the secondary mechanism is the effective magnetic softening of the shell magnetization in the PS and RS nanowires, which no longer acts as a collection of nucleation centers for magnetization reversal. 
The magnetic softening of the shell interface magnetization results in lower domain wall velocities in the core, as seen in Fig.~\ref{fig:velocities} and also by comparison of domain wall mobility values for the MS, PS and RS nanowires in Table~\ref{table:hchb}.
%----------------------------------
% FIG 7 - v(H)
%----------------------------------
The contribution of the shell-interface magnetization $\Mshif$ in the modification of the wall velocities and mobility can be also deduced from Fig.\ref{fig:velocities}. 
The RSf nanowire that has higher $\Mshif$ value than the MS nanowire (see Fig.~\ref{fig:M3_CryWidth}), exhibits a lower mobiltity. On a microscopic level, this trend is explained as the number of satisfied interface bonds is substantially higher for the RSf nanowire. These bonds being in their lowest energy state oppose their reversal under the applied field, acting as soft pinning centers for domain wall propagation. 

\begin{figure} [htb!]
\centering	\includegraphics[width=0.95\linewidth]{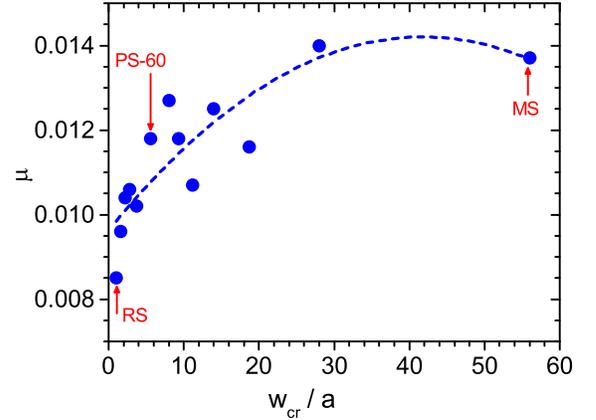}
\caption{
	(Color online) Dependence of DW mobility on shell grain size. Arrows indicate the structural models reported in Table~\ref{table:hchb}. 
	Dashed line is a guide to the eye. 
	Other parameters: $R_c=5a, \tsh=3a, L_c=50a$ and $T=0.01\Jfm$.
	}
\label{fig:Mob_CryWidth}
\end{figure}

Thus, drag of the AF interface moments  and increased $\Mshif$ magnitude due to polycrystallinity are the two factors acting in synergy to suppress the domain wall mobility in polycrystalline samples. A systematic decrease of wall mobility with decreasing grain size is shown in Fig.~\ref{fig:Mob_CryWidth}.
%----------------------------------
% FIG 8 - Mob_Crywidth
%----------------------------------

%================================================================
\subsection{Angular dependence of $\Hc$ and $\Heb$ }
\begin{figure} [htb!]
\centering
\includegraphics[width=0.95\linewidth]{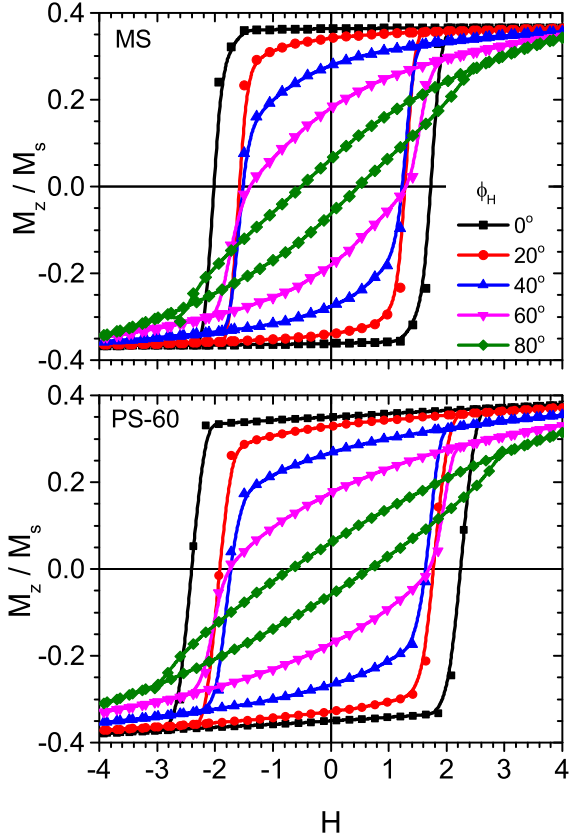}
\caption{
	(Color online) Angular dependence of hysteresis loops of nanowires with monocrystalline shell (upper panel) and polycrystalline shell (lower panel). Parameters as in Table~\ref{table:hchb}.
}
\label{fig:loops_Hangle}
\end{figure}

%================================================================
In the present section, we study the changes introduced in the hysteresis behavior of core-shell nanowires, when the cooling field and the reversing field both lie at an angle $\phi_H$ with respect to the nanowire axis. 
In Fig.~\ref{fig:loops_Hangle} we show the angular dependence of the hysteresis loop for samples with monocrystalline and polycrystalline shells.
%----------------------------------
% FIG 9 - loops_Hangle
%----------------------------------

\begin{figure} [htb!]
	\centering	
	\includegraphics[width=0.95\linewidth]{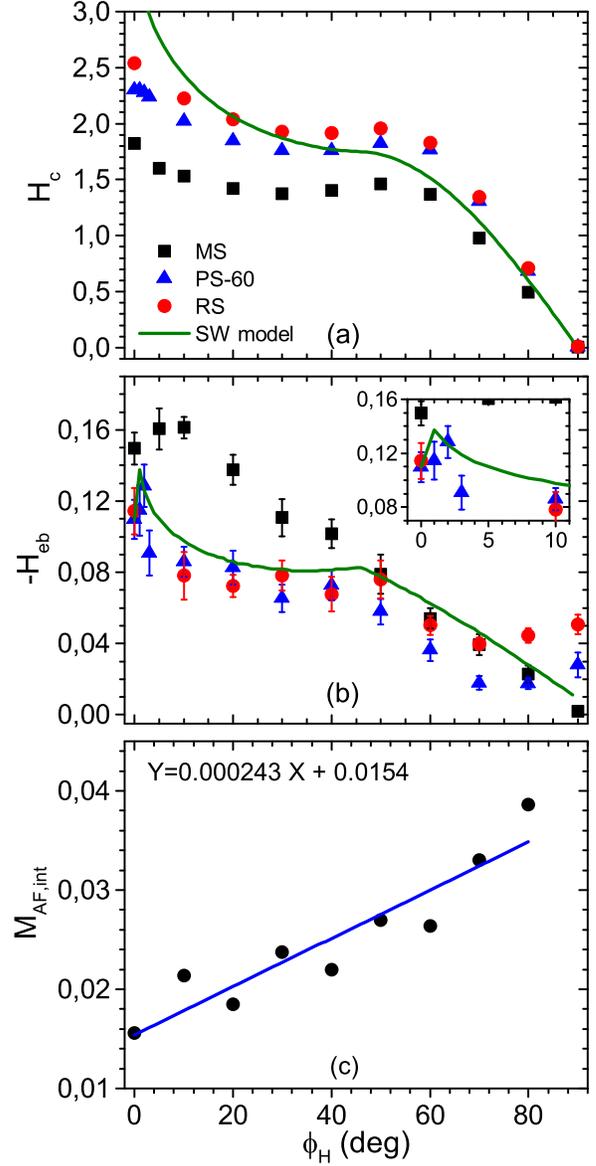}
	\caption{
		(Color online) Angular dependence of (a) coercivity and (b) exchange bias field for different structural models described in Table~\ref{table:hchb}. 
		The solid line is a Stoner-Wohlfarth model fitted to the simulated data with parameters $k=3.5, k_R=0.2$ and  $h_b=0.11$ and $\phi_b=5^\circ$.
		Error bars, when not indicated, are of the size of the marks. 
		Inset to (b) is the enlarged angular region around the sharp peak of $\Heb$ for the polycrystalline sample.
		(c) Linear dependence of the AF-interface magnetization ($M_{AF,int}$) of the monocrystalline nanowire (MS) on the angle of the applied field.
}
\label{fig:HcHeb_vs_Hangle}
\end{figure}

As the field angle relative to the nanowire axis increases, a gradual shrinking and tilting of the loop is observed, as both the crystallographic and shape anisotropies lie along the nanowire axis.  
Shell granularity does not change this trend.
Analysis of the hysteresis loops at different field angles provides the angular dependence of the coercivity and EB shown in Fig.~\ref{fig:HcHeb_vs_Hangle}.
A monotonous decrease of $\Hc$ with field angle is seen for monocrystalline and granular samples. 
However, the polycrystalline and random samples exhibit higher coercivity values  than the monocrystalline sample at all angles. This behavior is due to AF drag effect, discussed previously (see Fig.~\ref{fig:M3_CryWidth}).
%----------------------------------
% FIG 10 - Hc,Heb_vs_Hangle
%----------------------------------

A more complex behavior is seen in Fig.~\ref{fig:HcHeb_vs_Hangle}b in the angular dependence of $H_{eb}$. 
In particular, the monocrystalline nanowire exhibits a weak increase of $\Heb$ at small angles with a broad peak around $\phi_b\approx 10^\circ$.
This peak is attributed to the competition between two distinct factors, namely, the susceptibility of the AF shell and the projection of the bias field on the applied field direction.
First, we mention that the AF shell of the MS nanowire is a hard antiferromagnet with anisotropy along the z-axis. 
So, the susceptibility of the shell increases when the field angle changes from the parallel to the normal direction and consequently, the shell interface magnetization ($\Mshif$) increases too when the cooling field rotates with respect to the z-axis. 
Indeed, a linear fit $\Mshif \sim M_0 + \lambda \phi_b$, is seen in Fig.~\ref{fig:HcHeb_vs_Hangle}c.
Second, for fixed shell interface magnetization, the bias field decreases as $\Heb \sim cos\phi_H$ according to the Meiklejohn-Bean description of the EB effect\cite{nog99}.
Thus, an overall dependence $\Heb \sim (M_0 + \lambda\phi_H) \cdot cos(\phi_H)$ is expected for the MS nanowire, which is characterized by an optimum field angle for maximum $\Heb$ values. 
The optimum angle is $\phi_{max} \sim 10^\circ$ within our model for the MS nanowire.
When shell granularity is present, as in PS and RS nanowires, the shell becomes magnetically softer and the $\Heb$ values drop.
Furthermore, due to the random distribution of the easy axes of the shell grains, the susceptibility of the AF shell becomes isotropic and the broad peak of $\Heb$ at $\sim 10^\circ$ vanishes.  
However, a sharp but weak increase of $\Heb$ is seen for the polycrystalline sample for slightly off-axis directions. 
This behavior, despite being weak within our model is well understood within an effective SW model to be discussed next. 

%================================================================
 \subsection{Effective Stoner Wohlfarth model}
%================================================================
In this section, we fit our simulation results to an effective SW model.
We define the SW model in dimensionless units, as follows:
%----------------------------------
% EQN - SW model
%----------------------------------
\begin{eqnarray}
e(\phi) =
-\frac{1}{2}\cdot k \cdot cos^2(\phi)
-h \cdot cos(\phi-\phi_H)      \nonumber \\
-h_b \cdot cos(\phi-\phi_b)
-\frac{1}{2} \cdot k_R \cdot cos^2(\phi-\phi_H)
\label{eq:s-w}
\end{eqnarray}
where $k$ is the anisotropy of the FM assumed along the z-axis, $h$ is the applied field at an angle $\phi_H$, $h_b$ the bias field at an angle $\phi_b$ and $k_R$ is an effective anisotropy that accounts for the coercivity enhancement due to drag of the AF interface moments\cite{jim09}.
We obtain the values of the dimensionless parameters in Eq.~\ref{eq:s-w} from the simulation results (Fig.~\ref{fig:HcHeb_vs_Hangle}), as follows\cite{jim09}:
%$ k = H_c^{(mc)}/K_{FM}^{(mc)} \approx 2.4$,
$ k_R = \frac{1}{2} H_c^{(mc)}/K_{FM}^{(mc)} -1 \approx 0.2$
and
$h_b = \Heb^{(mc)} \approx 0.11$,
where the superscript $(mc)$ denotes values related to our Monte Carlo simulations. The anisotropy $k$ is not fitted to the coercivity value\cite{jim09}, but is treated as an adjustable parameter.  
This is because the coherent rotation mechanism, inherent to the SW model, overestimates the coercivity for an applied  field along the easy axis. The angle $\phi_b$ is also treated as an adjustable parameter.
We perform numerical minimization of Eq.~\ref{eq:s-w}, leading for each $h$ value to one (reversible part) or two (irreversible part) extreme points that correspond to the equilibrium magnetization value(s). 
In Fig.~\ref{fig:HcHeb_vs_Hangle}a,b we compare the results of the SW model for the angular dependence of $\Hc$ and $\Heb$ with the Monte Carlo data. 

We find that the Monte Carlo simulation data for the angular dependence of $\Heb$  and $\Hc$ can be satisfactorily described by a SW model with an effective off-axis unidirectional anisotropy making an angle $\phi_b \sim 5^\circ$ with the nanowire axis. 
As previously, reported in the case of FM/Af bilayers, an effective off-axis anisotropy arises from the frustration of the AF interface moments due to interface roughness \cite{jim09}. 
In the case of core-shell nanowires studied here, the source of magnetic frustration is the anisotropy disorder occurring at the core-shell interface due to shell granularity. 

As a final remark, we discuss briefly recent experiments on electrodeposited Co/CoO nanowires\cite{gan17}, which showed giant exchange bias and an effective off-axis unidirectional anisotropy at a large angle ($\phi_b \sim 30^\circ$) leading to an increase of the bias field by $ \sim 20\%$ when the cooling field is applied at $\phi_H\sim 30^\circ$ relative to the wire axis. 
Our simulations for Co/CoO nanowires, reproduce a similar trend, namely the appearance of an off-axis ($\phi_b \sim 5^\circ$) anisotropy and a weak increase of the bias field ($\sim 10\%$) for a field applied at $\phi_H\sim 1^\circ$ relative to the nanowire axis.
Most importantly, the simulations point to the shell polycrystallinity as a source of generating effective off-axis anisotropy and optimum field directions for maximizing the EB effect in magnetic core-shell nanowires. 
However, the quantitative discrepancy between our simulations results and the experimental findings\cite{gan17} we believe that is caused by the relative stronger bias effect in these samples, which most probably stems from atomic scale details of the interface structure leading to strong uncompensation, as for example, crystallographic orientation, interface roughness, alloying, etc, which are not considered in the present model.

%================================================================
\section{Summary}
%================================================================
We have studied the isothermal magnetic hysteresis of cylindrical Co/CoO nanowires with core-shell morphology using a classical spin model and Monte Carlo simulations.
We  have demonstrated that polycrystallinity of the CoO shell causes enhancement of coercivity  and suppression of exchange-bias field relative to the monocrystalline shell. 
This behavior is mainly attributed to enhanced drag of the AF interface spins during reversal of the core spins. 
The same mechanism is responsible for reduction of domain wall mobility. 
Additionally, Co/CoO nanowires with monocrystalline shell, exhibit a weak ($\approx 10\%$) increase of their bias field when the reversing field is applied at a small angle ($\approx 10^\circ$) relative to the nanowire axis. 
On the other hand, shell granularity introduces an effective off-axis ($\phi_b \sim 1^\circ$) unidirectional anisotropy due to frustration of the shell interface spins, leading to weak increase ($\sim 10\%$) of the bias field in an off-axis direction.
Our results are in qualitative agreement with recent experimental observations in Co/CoO nanowires\cite{gan17}, and point to shell polycrystallinity as a source of generating off-axis unidirectional anisotropy in FM core - AF shell nanowires and exchange bias enhancement in off-axis directions. 

%================================================================
\section*{Acknowledgments}
%================================================================
DK acknowledges the hospitality in the Physics Department (University of Athens) during the course of this work and the financial support for dissemination of the work by the School of Pedagogical and Technological Education through project \textit{Strengthening of Research in ASPETE}.
%================================================================
\section*{References}
%================================================================
%\bibliography{ref_exchange_bias_v3}

\end{document}